\documentclass[10pt,conference]{IEEEtran}
\IEEEoverridecommandlockouts

\renewcommand{\arraystretch}{0.90}  

\usepackage{cite}
\usepackage{amsmath,amssymb,amsfonts}
\usepackage{algorithm}
\usepackage{algorithmic}
\usepackage{graphicx}
\usepackage{booktabs}
\usepackage{listings}
\usepackage{enumitem}
\usepackage{multirow}
\usepackage{placeins}
\usepackage{subcaption}
\usepackage{tabularx}
\let\origtextunderscore\textunderscore
\renewcommand{\textunderscore}{\origtextunderscore\discretionary{}{}{}}
\usepackage{tcolorbox}
\tcbuselibrary{skins}
\usepackage{xspace}
\usepackage{pifont}
\usepackage[table]{xcolor}  

\usepackage{tikz}
\usetikzlibrary{arrows.meta, positioning, shapes.geometric, fit, backgrounds, calc}
\usepackage{hyperref}
\usepackage{url}
\usepackage{textcomp}

\providecommand{\citep}[1]{\cite{#1}}
\providecommand{\citet}[1]{\cite{#1}}

\providecommand{\Description}[1]{}

\setlist{nosep, leftmargin=*}

\definecolor{codebg}{gray}{0.95}
\newcommand{\tool}{\textsc{TraceProbe}\xspace}
\newcommand{\converge}{\textsc{Converge}\xspace}
\newcommand{\insight}{\textsc{Insight}\xspace}

\newcommand{\eg}{e.g.,\xspace}

\hypersetup{
  colorlinks=true,
  linkcolor=black,
  citecolor=black,
  urlcolor=blue!60!black,
  breaklinks=true,
}

\begin{document}

\title{What Resolve Rate Hides: Trajectory Structure Diagnostics for Coding Agents}

\author{%
\IEEEauthorblockN{%
Rui Shu, Chun Yong Chong, Xin Zhou, Yun Peng, Zihan Wu\\
Xu Han, Zeyang Zhuang, Guowen Yuan, Yuan Wang}%
\thanks{Preprint. Under review.}%
}

\maketitle

\begin{abstract}
Coding agents are ranked almost entirely by \emph{resolve rate}: whether their final patch passes the target tests.
Yet two agents can reach the same outcome through very different processes, and a single pass/fail label says nothing about why a run failed or why an accepted run spent extra steps, time, or tokens.
This process evidence lives in the trajectory, which records a run's searches, reads, edits, tool calls, validation, and reversions.
However, raw traces are heterogeneous and hard to compare across runs.
We present \tool, a trajectory-diagnostic framework that recovers what resolve rate hides.
\tool normalizes each raw run into a canonical nine-type action taxonomy with deterministic effect labels, then applies two rule-based modules: \insight names single-trajectory anti-patterns adapted from established debugging practice (\eg search loops, verification skips), while \converge aligns pairs of runs and classifies where their behavior diverges under controlled references.
Applying \tool to 2{,}500 trajectories from five production settings on SWE-Bench Verified, we find that (i) file choice is too coarse to separate success from failure, whereas function selection and completion behavior localize it; (ii) \insight anti-patterns act mainly as corpus-level difficulty clues, with search loops the most stable; and (iii) even resolved runs differ in how quickly they reach relevant code and how much failed work they incur.
Trajectory structure thus adds auditable diagnostic context to outcomes by localizing inspection targets, suggesting failure hypotheses, and prioritizing runs for review.

\end{abstract}


\section{Introduction}
\label{sec:intro}

LLM-powered coding agents are now widely used in developer environments~\citep{Anthropic2025ClaudeCode,OpenAI2025Codex,OpenCode2025}.
They inspect repositories, edit files, invoke tools, run tests, and iterate on feedback.
As these systems move into production workflows, practitioners need to compare not only whether an agent solved a task, but also how it behaved while attempting the task.
As the key metric, \emph{resolve rate} records whether the final patch passes target tests.
This outcome is useful for ranking systems and guiding release decisions, but it hides the process differences that matter to agent developers.

\begin{figure}[t]
\centering
\includegraphics[width=\columnwidth]{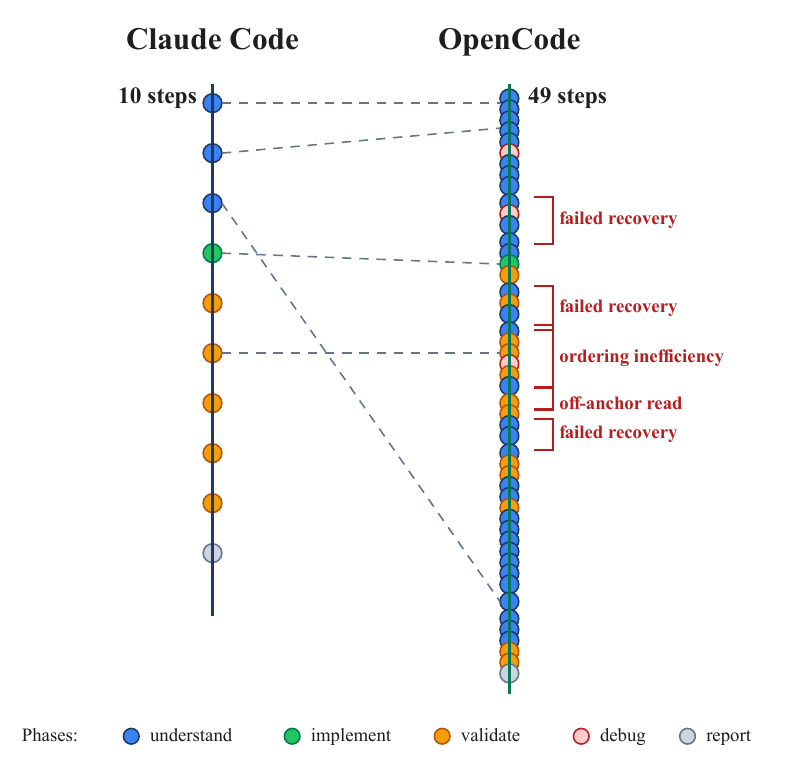}
\caption{In this example, Claude Code (Opus~4.6) and OpenCode (GLM-5) both \texttt{RESOLVE} the task (SWE-Bench \texttt{pytest-7982}), but their trajectories differ substantially. Claude Code reaches a targeted fix in 10 steps with no failed actions, while OpenCode takes 49 steps and includes repeated failed/recovery spans. Colored dots show workflow phases, and dashed gray lines show actions aligned by our framework. Red brackets provide informal visual annotations for this example, and our analysis later defines the patterns and divergence signals used for measurement.}
\label{fig:motivating}
\end{figure}

This limitation appears even when two runs reach the same outcome.
Figure~\ref{fig:motivating} shows an example where Claude Code (Opus~4.6) and OpenCode (GLM-5) both resolve the same SWE-Bench task.
The pass/fail label treats both runs as successes, but their processes differ.
Claude Code reaches a targeted fix with no failed actions, while OpenCode takes a longer path with repeated failures and recovery work.
This example illustrates why process evidence is required in addition to pass/fail outcomes.
Prior work has approached this need from two directions.
One direction questions whether pass/fail outcomes are reliable, showing that generated patches can exploit weak tests, recall benchmark artifacts, or be affected by infrastructure and run-to-run noise~\citep{DBLP:conf/acl/YuZHK25,DBLP:journals/corr/abs-2503-15223,DBLP:journals/corr/abs-2506-12286,Anthropic2026InfrastructureNoise,OpenAI2025SWEBenchRetirement}.
The other direction evaluates or characterizes trajectories directly. AgentBoard scores progress against predefined sub-goals~\citep{DBLP:conf/nips/MaZZYYJLKH24}, TrajEval grounds failure analysis against reference patches~\citep{DBLP:journals/corr/abs-2603-24631}, process-centric analysis profiles behavior~\citep{DBLP:journals/corr/abs-2512-02393}, and Majgaonkar et al.~\cite{DBLP:journals/corr/abs-2511-00197} summarize success and failure trajectories with aggregate statistics.
These methods show that agent trajectories contain useful evidence.
However, they still leave a gap for process diagnosis.
Practitioners need deterministic diagnostics that name concrete trajectory patterns, compare runs under controlled references, and state the oracle assumptions behind each signal.

Closing this gap is not trivial because raw trajectories are not directly comparable.
A single run interleaves multiple heterogeneous events with no shared schema, including reasoning, tool calls, edits, shell output, and test logs.
The relevant signals are also sparse, such as an early read of the wrong file, a skipped validation step, or repeated searches for the same evidence.
Aggregate counts lose this local context, while manual inspection does not scale across large trajectory corpora.
Process diagnosis therefore requires a representation that preserves local evidence and still supports corpus-level comparison.
On top of this representation, the analysis should support comparison across agent settings, not only case-by-case debugging.
It should give practitioners concrete failure clues, point to avoidable effort in resolved runs, and compare agent behavior under controlled references.
To be auditable, these signals need deterministic rules, explicit oracle assumptions, and named structural patterns rather than LLM judgments or aggregate statistics alone.

This work asks three diagnostic questions: \textit{what observable clues are associated with failure, where successful runs spend avoidable effort, and what comparisons among different agents' runs reveal about behavioral divergences}.
We present a framework named \textbf{\tool} to answer these questions through a shared representation for within-trajectory diagnosis and between-trajectory comparison.
\tool normalizes each raw trajectory into canonical actions from a nine-type taxonomy: file read, file write, search, command, sub-agent spawn, plan, navigate, fetch, and reason.
It then assigns each action a deterministic effect label from observable state transitions.
Two rule-based modules build on this substrate.
The \insight module profiles one trajectory for recurring failure modes such as \emph{search loops} and \emph{verification skips}, while the \converge module aligns two trajectories and classifies divergences such as \emph{off-anchor exploration} and \emph{rapid rewrite}.
The structural detectors and most effect labels are oracle-free, while off-anchor labels, milestones, and benchmark-informed divergence classifications use gold-patch anchors when available.

We scope \tool as a diagnostic framework, not a replacement for resolve rate, a per-run root-cause oracle, or a learned policy for fixing agents.
It provides auditable signals for comparing process distributions across agent settings, especially when the study design holds the task, scaffold, or model fixed.
Within this scope, we apply \tool to 2,500 trajectories from SWE-Bench Verified tasks across five production settings spanning three scaffolds (Claude Code, Codex, OpenCode) and three model backbones (Opus~4.6, GPT-5.4, GLM-5).
This design supports both same scaffold model contrasts and same model scaffold contrasts.
Our study surfaces three results.
First, comparing failed and successful runs on the same task shows where failure evidence should be localized.
File choice alone is too coarse, because two successful runs can touch the same files, but clearer signals appear in function selection and completion behavior.
Second, the \insight module mainly provides corpus-level difficulty clues.
Search loops are the most stable clue, while other signals are split-sensitive.
A smaller completion evidence signal shows only a modest failed-run gap.
Third, the \converge module milestones and classifications show that successful runs can differ in how quickly they reach relevant code, how much failed work they perform, and how their process changes with the scaffold or model.

The main contributions of this paper are threefold:
\begin{enumerate}
\item We introduce a deterministic canonical action representation for coding agent trajectories that separates oracle-free structural diagnostics from anchor-grounded milestones and divergence evidence.
\item We define a detector and alignment methodology that turns established SE and debugging anti-pattern concepts into exact trajectory predicates and produces reference-scoped divergence evidence with explicit oracle assumptions.
\item We conduct a multi-setting study showing that trajectories reveal differences between agent settings that resolve rate hides, including where runs diverge, when they reach relevant code, and how much failed work they perform. All data and code are released at \url{https://doi.org/10.5281/zenodo.20789918} for future study.
\end{enumerate}

\section{Related Work}
\label{sec:related}

\subsection{Code Agent Evaluation}

SWE-Bench~\citep{DBLP:conf/iclr/JimenezYWYPPN24} uses resolve rate as the default metric for coding agents, but a final pass/fail outcome is only a coarse signal.
Prior work shows two reasons for caution.
First, a generated patch can satisfy permissive tests while missing developer intent~\citep{DBLP:conf/acl/YuZHK25,DBLP:journals/corr/abs-2503-15223}.
Second, reported scores can be distorted by benchmark memory, infrastructure noise, benchmark saturation, or run-to-run variance~\citep{DBLP:journals/corr/abs-2506-12286,Anthropic2026InfrastructureNoise,OpenAI2025SWEBenchRetirement,DBLP:journals/corr/abs-2602-07150}.
Recent benchmarks respond by making outcome evaluation broader and more robust.
Some benchmarks collect live or contamination-resistant tasks~\citep{DBLP:conf/nips/ZhangHZKLXWWHFN25,DBLP:journals/corr/abs-2505-20411}, while others target enterprise repairs, multilingual repositories, unified coding agent skills, or security sensitive fixes~\citep{DBLP:journals/corr/abs-2509-16941,DBLP:journals/corr/abs-2504-02605,DBLP:journals/corr/abs-2504-08703,DBLP:journals/corr/abs-2511-05459,DBLP:journals/corr/abs-2509-22097}.
These efforts make final outcomes more informative.
Our work asks a complementary question: \textit{what process produced that outcome?}
Unlike AgentBoard~\citep{DBLP:conf/nips/MaZZYYJLKH24}, which scores progress against predefined sub-goal sequences, \tool normalizes observed agent events and compares trajectories directly without requiring a hidden ideal path.
Agent scaffolding systems such as SWE-agent, OpenHands, and Agentless~\citep{DBLP:conf/nips/YangJWLYNP24,DBLP:conf/iclr/0001LSXTZPSLSTL25,DBLP:journals/corr/abs-2407-01489} improve coding-agent performance, but their evaluations still focus mainly on final outcomes.
\tool adds a complementary process view (i.e., a common representation) for asking how agents searched, edited, validated, recovered, and diverged.

\subsection{Agent Trajectory Analysis}

Beyond final-outcome benchmarks, a growing body of work evaluates the trajectories that agents produce.
TrajEval~\citep{DBLP:journals/corr/abs-2603-24631}, TRAJECT-Bench~\citep{DBLP:journals/corr/abs-2510-04550}, and process-level configuration evaluation~\citep{DBLP:journals/corr/abs-2510-25694} support fine-grained trajectory assessment using reference patches, synthetic reference trajectories, or environment configuration tasks.
Process-centric analysis~\citep{DBLP:journals/corr/abs-2512-02393} and empirical trajectory characterization~\citep{DBLP:journals/corr/abs-2511-00197,DBLP:journals/corr/abs-2604-02547,DBLP:conf/kbse/BouzeniaP25,DBLP:journals/corr/abs-2506-08311} also show that trajectory structure matters.
These studies show that trajectories are useful, but they usually summarize them as scores, graphs, profiles, or correlations.
Our work instead pairs oracle-free detectors with run-to-run alignment.
It maps each raw trace to a shared action format, then applies named detector rules with fixed thresholds.
As a result, each measurement is tied to an explicit rule and can be compared across agent settings.
\tool targets a different use case than trajectory scoring or LLM judge localization~\citep{DBLP:journals/corr/abs-2505-08638}, i.e., corpus level comparison of how agent settings behave.
For that purpose, \converge adapts event sequence matching and visual comparison ideas~\citep{DBLP:journals/tvcg/GuoJCGZC22} to coding agent traces: it aligns two runs, localizes unmatched or reordered action spans, and reports reference-scoped diagnostics with fixed thresholds.

Trajectory analysis is also used for failure attribution, inspection, monitoring, and agent improvement.
Failure attribution systems~\citep{DBLP:journals/corr/abs-2602-02475,DBLP:journals/corr/abs-2605-26563,DBLP:conf/emnlp/OuGGNY25,DBLP:conf/icml/ZhangY0LHZL0W0W25,DBLP:journals/corr/abs-2509-03312,DBLP:journals/corr/abs-2509-25370} localize why failed trajectories went wrong, while MAST and silent failure detectors focus on multi-agent failures~\citep{DBLP:conf/nips/CemriPYACTKPKRZ25,DBLP:journals/corr/abs-2511-04032}.
Our scope is single-agent SWE coding tasks. \tool compares successful and failed runs and links each diagnostic signal to trace evidence.
Adaptive monitors such as TRACE~\citep{mittapalli2026tracetrajectoryreasoningadaptive} trade repeatability for model-based flexibility, while our work uses fixed rules for repeatable measurement.
Because these systems answer different questions, direct accuracy comparisons would not measure the same construct. We therefore use baselines defined on the same trajectories: success/failure outcomes alone and raw telemetry such as steps, tokens, and tool counts.
Inspection and visualization systems help users examine traces directly~\citep{DBLP:conf/aaai/DesmondLIJSMP25,DBLP:journals/corr/abs-2504-08696,DBLP:journals/corr/abs-2508-12555,DBLP:journals/corr/abs-2604-11641}.
Other work uses trajectories to improve agents during or after execution, for example through reward models, course correction, trajectory reuse, harness evolution, or richer trace representations~\citep{DBLP:conf/www/XiLLZCWJZGWJGZH26,DBLP:journals/corr/abs-2505-20732,DBLP:journals/corr/abs-2509-02360,DBLP:conf/iclr/XuLSWWMX025,DBLP:journals/corr/abs-2603-25158,DBLP:journals/corr/abs-2603-10600,cuadron2026shepherd,DBLP:journals/corr/abs-2604-25850}.
\tool's core contribution is a reusable detector catalog for real SWE trajectories, which helps practitioners apply the same diagnostic rules across agent settings and compare process differences with trace level evidence.

\subsection{Anti-Pattern Detection in Software Engineering}
\label{sec:rw-antipatterns}

Software engineering research has long used named patterns to describe code smells~\citep{DBLP:books/daglib/0019908}, design anti-patterns~\citep{brown1998refactoring}, organizational debt~\citep{DBLP:journals/jisa/TamburriKLV15}, socio-technical anti-patterns in ML-enabled software~\citep{DBLP:conf/icse/MailachS23}, and CI/CD configuration issues~\citep{DBLP:journals/ese/ZampettiVPCGP20}.
Many of those settings scan static artifacts, but agent trajectories are dynamic event sequences whose shape depends on scaffold, prompt, and tool interface.
Our work brings this anti-pattern tradition to trajectory analysis by using named signals for recurring agent behavior.
It maps raw events to canonical actions with deterministic effect labels, then defines structural detectors over that representation.
For example, analysis paralysis and thrashing~\citep{brown1998refactoring} motivate search-loop and re-read-churn detectors.
Prior trajectory characterization and failure taxonomy studies identify important behavior patterns~\citep{DBLP:journals/corr/abs-2509-13941}.
\tool turns such patterns into a detector catalog that can be tested by outcome group, reused across agent settings, and used to explain what changed in an agent's process.

\section{\tool: Offline Trajectory Analysis}
\label{sec:approach}

\newcommand{\phIntake}{\textcolor{blue!70!black}{\textbf{(Intake)}}}
\newcommand{\phUnderstand}{\textcolor{teal!70!black}{\textbf{(Understand)}}}
\newcommand{\phPlan}{\textcolor{orange!75!black}{\textbf{(Plan)}}}
\newcommand{\phImplement}{\textcolor{green!50!black}{\textbf{(Implement)}}}
\newcommand{\phValidate}{\textcolor{purple!75!black}{\textbf{(Validate)}}}
\newcommand{\phDebug}{\textcolor{red!75!black}{\textbf{(Debug)}}}
\newcommand{\phReport}{\textcolor{brown!80!black}{\textbf{(Report)}}}
\newcommand{\phCross}{\textcolor{gray!60!black}{\textbf{(Cross-cutting)}}}

\begin{figure*}[!t]
\centering
\includegraphics[width=0.95\textwidth]{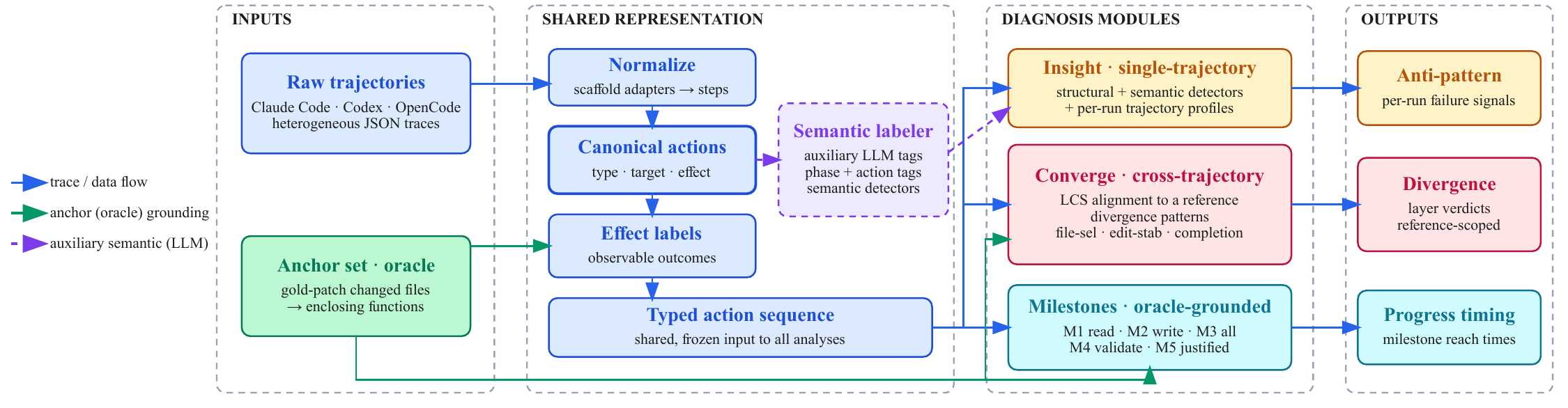}
\Description{Architecture diagram with three left-to-right stages. The input panel shows raw Claude Code, Codex, and OpenCode trajectory records and a gold-patch anchor set used as an oracle. The TraceProbe core panel normalizes trajectories, maps them into a unified step schema, converts steps into canonical actions, and assigns deterministic effect labels. The analysis panel has three parallel modules built on the shared representation: Insight for single-trajectory anti-pattern detection and trajectory profiles, Converge for cross-trajectory alignment, divergence classification, and layer classifications for file selection, edit stability, and completion behavior, and Milestones for single-trajectory oracle-grounded progress timing.}
\caption{\tool architecture. Raw trajectories are normalized into a shared canonical representation, then analyzed by \insight for single-run patterns and by \converge for comparison against another run chosen as the reference trajectory.}
\label{fig:overview}
\end{figure*}

\begin{table*}[!t]
\centering
\setlength{\tabcolsep}{3pt}
\caption{Step-level label schema in \tool. Deterministic canonical actions, effect labels, and workflow phases support the main analyses, while optional LLM semantic tags are used only by detectors marked semantic.}
\label{tab:step-labels}
\renewcommand{\tabularxcolumn}[1]{m{#1}}%
\begin{tabularx}{0.95\textwidth}{|>{\centering\arraybackslash}m{2.3cm}|>{\scriptsize\ttfamily\raggedright\arraybackslash}X|>{\raggedright\arraybackslash}m{4.4cm}|}
\hline
\multicolumn{1}{|c|}{\rule[0.0pt]{0pt}{10pt}\textbf{Layer}} & \multicolumn{1}{c|}{\textbf{Label Set}} & \multicolumn{1}{c|}{\textbf{Role in Analysis}} \\
\hline
\rowcolor{gray!15}\multicolumn{3}{|l|}{\rule[0.0pt]{0pt}{10pt}\textit{Deterministic layers}}\\
\hline
\rule[-6.5pt]{0pt}{18pt}Canonical Action &
FILE~READ, FILE~WRITE, SEARCH, COMMAND, PLAN, NAVIGATE, FETCH, AGENT~SPAWN, REASON &
Deterministic event sequence used by \insight, \converge, and process-profile summaries. \\
\hline
\rule[-6.5pt]{0pt}{18pt}Effect Label &
SURVIVED, FAILED, REVERTED, JUSTIFIED, RECORDED, OFF-ANCHOR, REASONING &
Observable action outcome used in detector predicates and process-cost summaries. \\
\hline
\rule[-6.5pt]{0pt}{18pt}Workflow Phase &
UNDERSTAND, PLAN, IMPLEMENT, DEBUG, VALIDATE, REPORT &
Coarse workflow stage used for phase-level profiles. Semantic tags use the same phase set. \\
\hline
\rowcolor{gray!15}\multicolumn{3}{|l|}{\rule[0.0pt]{0pt}{10pt}\textit{Semantic layer: LLM-based fine-grained action tags, grouped by their phase}}\\
\hline
Understand &
spec~intake, scope~clarification, constraint~extraction, tool~discovery, delegated~research, external~research, file~discovery, code~reading, context~synthesis &
\multirow{6}{4.4cm}[-29pt]{\raggedright Fine-grained tags used only by detectors marked semantic.} \\
\cline{1-2}
Plan &
plan~management, task~breakdown, sequencing, dependency~mapping, risk~planning & \\
\cline{1-2}
Implement &
implement~api~schema, implement~runtime~logic, implement~tests, implement~generated~artifacts, implement~config, implement~docs, implement~refactor, implement~migration & \\
\cline{1-2}
Debug &
debug~reproduction, debug~root~cause, debug~hypothesis~test, debug~fix~selection & \\
\cline{1-2}
Validate &
validation~run, validation~review, validation~lint, validation~build, validation~unit~tests, validation~integration~tests, validation~e2e~tests, validation~regression, validation~performance & \\
\cline{1-2}
Report &
progress~update, final~reporting, change~summary, blockers~risks, next~steps~handoff & \\
\hline
\end{tabularx}
\end{table*}

\subsection{Architecture Overview}

At a high level, Figure~\ref{fig:overview} shows how \tool separates representation construction from diagnostic analysis.
Before describing the modules, we define six terms used throughout the methodology.
A \emph{reference} is another trajectory chosen for comparison.
An \emph{anchor set} is the set of task-relevant files supplied by a benchmark patch when available.
A \emph{canonical action} is a typed event record, and an \emph{effect label} records its observable outcome.
A \emph{divergence span} is a contiguous block of actions that remains unmatched or reordered after \converge aligns two runs.
A \emph{milestone} is the first step that reaches a task-relevant artifact, and an \insight detector flags a named pattern in one trajectory.
\tool first converts raw trajectories into normalized steps, typed canonical actions, and deterministic effect labels.
\insight then analyzes one run without a reference or anchor set.
\converge compares a run with a reference trajectory and labels its divergence spans.
Milestones use the anchor set to record when a run first reaches task-relevant artifacts.

\subsection{Shared Representation}
\label{sec:shared-rep}
\label{sec:anchor-set}

To compare trajectories across scaffolds, \tool first builds a shared deterministic representation.
The representation starts with normalized steps that preserve event context in a common schema.
Each step is then converted into a canonical action, which records the action type and comparable target.
On top of this action sequence, \tool derives workflow phases to summarize coarse progress and effect labels to record observable action outcomes.

This process starts with scaffold adapters.
They translate scaffold-specific traces into a common step schema that preserves role, typed tool-call records, resource use, timing, and error status.
This layer absorbs format differences across agents.
For example, Claude Code may split one API response across multiple trajectory entries, while OpenCode stores comparable information inline.
\tool then maps each normalized step to a canonical action with an action type and target.
The action type uses the nine-label set in Table~\ref{tab:step-labels} to distinguish the kind of work performed by the step.
The target stores the comparable object for the action, such as a path, command, query, scope, or plan summary.
File targets use repo-relative paths, allowing \converge to align the same artifact across different local work directories.

\tool also derives a deterministic workflow phase from observed step fields.
The classifier assigns the phase from the step's dominant event, separating planning, implementation, validation, debugging, repository inspection, and reporting activity.
These phases support phase-level profiles and provide the phase set used by semantic tags.

To decide whether a read, search, or milestone touched task-relevant code, \tool needs a task-relevance set.
When a benchmark supplies a ground-truth patch, as in SWE-Bench Verified, the modified files define the trajectory's \emph{anchor set}, or gold-patch file set.
\tool uses these anchors for off-anchor labels, milestone extraction, and trajectory comparisons that refer to task-relevant artifacts.
An off-anchor label marks a successful read or search outside the anchor set.
For non-benchmark traces, \tool can instead derive task relevance from surviving writes and test/import references.
The detector catalog uses only trajectory events, not the SWE-Bench issue description.
As a result, same-task comparisons are driven by what agents did rather than by prompt text that is identical across agents.

After action typing, \tool assigns each action a deterministic effect label using only observed trajectory data, not an LLM judge.
A write is \emph{survived} when it persists to the final state, and \emph{reverted} when a later write to the same file overwrites it.
An action is \emph{justified} when it reads or searches a task-relevant file, or runs a validation command.
A non-workspace meta-action, such as a plan update, navigation, fetch, or sub-agent spawn, is \emph{recorded} when it succeeds.
When reporting effect percentages, \tool excludes reasoning-only steps because they do not change files or return command status.
These labels describe what happened during the run, not whether the action was correct.

File-level anchors can still be too coarse when two runs touch the same file but edit different code regions.
For function-level analysis, \tool deterministically maps each patch hunk to the most-specific enclosing \texttt{def} or \texttt{class}, using the hunk header and an indentation-aware scan without an LLM.
When the resolver cannot identify an enclosing function, the hunk remains a valid file-level anchor.

The semantic layer is separate from this deterministic core.
For detectors marked semantic, the step labeler prompts an LLM to choose one phase/action pair from the closed taxonomy in Table~\ref{tab:step-labels}.
The parser validates each response and derives the phase from a valid action when needed.
Because these tags are produced by an LLM, they may change if the labeling model, prompt, or provider behavior changes.
We therefore use them only for detectors marked semantic and for exploratory analyses, not for the main deterministic claims.

\subsection{Anti-Patterns: Single-Trajectory Diagnosis}
\label{sec:antipatterns}

In \insight, an anti-pattern detector is a deterministic predicate over one trajectory.
If the predicate is satisfied, \insight reports a named pattern together with the local steps that support it.
To keep this evidence auditable, the reported catalog is intentionally selective rather than exhaustive.
A detector is included only when its triggering condition can be stated exactly, its inputs are observable tool-event features, and its numeric threshold is frozen before evaluation.

Table~\ref{tab:detector-catalog} gives the exact predicates, thresholds, and inputs that support claims in this work.
The first block contains the single-trajectory detectors used by \insight.
For traceability, the same table also lists the frozen cross-trajectory patterns and milestone definitions introduced in the next subsections.
Structural detectors support the main single-trajectory claims, while semantic detectors remain exploratory.
Ambiguous patterns such as rewrites move to cross-trajectory comparison, where a reference can disambiguate them.

\begin{table*}[!t]
\centering
\caption{Measurement catalog used in the reported analyses. Each row gives a named detector, its frozen predicate, the required inputs, and how the signal is used in the results. Thresholds are fixed before the stability, sensitivity, and transfer checks.}
\label{tab:detector-catalog}
\scriptsize
\setlength{\tabcolsep}{3pt}
\setlength{\extrarowheight}{0pt}
\renewcommand{\tabularxcolumn}[1]{m{#1}}%
\begin{tabularx}{0.95\textwidth}{|>{\centering\arraybackslash}m{2.05cm}|>{\raggedright\arraybackslash}X|>{\raggedright\arraybackslash}m{3.3cm}|>{\raggedright\arraybackslash}m{3.0cm}|}
\hline
\multicolumn{1}{|c|}{\rule[0.0pt]{0pt}{10pt}\textbf{Detector}} & \multicolumn{1}{c|}{\textbf{Frozen Predicate}} & \multicolumn{1}{c|}{\textbf{Input and Condition}} & \multicolumn{1}{c|}{\textbf{Use in Results}} \\
\hline
\rowcolor{gray!15}\multicolumn{4}{|l|}{\rule[0.0pt]{0pt}{10pt}\textit{Structural single-trajectory detectors (deterministic; no semantic labeler)}}\\
\hline
Search Loop & At least 10 consecutive SEARCH or FILE\_READ actions with no FILE\_WRITE and no validation COMMAND between them. & Canonical action types and command classifier; no oracle. & Primary failure-associated clue; stable under post-freeze checks. \\
\hline
Re-read Churn & Same canonical file path is read at least 3 times within a 10-action window, with no intervening write to that file. & Canonical FILE\_READ/FILE\_WRITE targets; no oracle. & Secondary failure-associated clue; split-sensitive. \\
\hline
Tool Oscillation & For one file, at least 2 READ--WRITE--READ cycles where the middle write is deterministically labeled failed or reverted. & Canonical targets plus effect labels; no oracle. & Low-prevalence process clue; not task-fixed. \\
\hline
No Formal Tail Validation & No recognized validation command in the overlap of the post-last-source-write region and final 5 actions. & Command classifier; no oracle; descriptive late-run signal. & Late-run behavior control; not treated as failure evidence. \\
\hline
Unsupported Completion Claim & Final reasoning text matches a word-boundary fixed/done/resolved regex and no successful validation command appears after the last source write. & Final REASON text plus command outcomes; no oracle. & Within-task completion-evidence signal; does not judge statement truth. \\
\hline
Shell-over-Tool & A shell command's first word is cat/head/tail/less/more, grep-family, rg/ag, or find while structured read/search tools are exposed. & Requires observable shell and tool surface; no oracle. & Scaffold behavior control; not treated as aggregate failure evidence. \\
\hline
Structured Plan Absence & At least 5 FILE\_WRITE actions occur with no preceding structured-plan or TODO tool call (tool-gated; weaker variant). & Canonical FILE\_WRITE actions plus plan-tool surface; no oracle. & Plan-phase failure-associated clue. \\
\hline
Redundant Search & Same exact-normalized SEARCH query recurs at least 2 times within a 10-action window. & Canonical SEARCH query text; no oracle. & Cross-cutting failure-associated clue; distinct from search loop. \\
\hline
\rowcolor{gray!15}\multicolumn{4}{|l|}{\rule[0.0pt]{0pt}{10pt}\textit{Semantic single-trajectory detectors (labeler-based; require phase/action labels)}}\\
\hline
Phase Oscillation & At least 3 transitions between the same two semantic phases within a 6-action window. & Requires semantic phase labels (optional labeler); no oracle. & Semantic process clue; failure-associated. \\
\hline
Semantic Fruitless Exploration & At least 5 code-read actions where at least 4 read files never appear in any later implement step. & Requires semantic phase/action labels (optional labeler); no oracle. & Understand-phase failure-associated clue. \\
\hline
\rowcolor{gray!15}\multicolumn{4}{|l|}{\rule[0.0pt]{0pt}{10pt}\textit{Cross-trajectory divergence patterns (deterministic; reference-aligned)}}\\
\hline
Off-anchor Exploration & Compared READ/SEARCH targets have off-anchor ratio at least 0.85 against the reference read set or, when available, the gold changed-file set. & Cross-trajectory alignment plus reference or anchor set. & File-selection divergence; anchor-grounded when gold files are used. \\
\hline
Scope Drift & Compared writes target files outside the reference write set or, in benchmark-informed mode, outside the gold changed-file set. & Cross-trajectory write targets plus reference/anchor set. & File-selection classification evidence. \\
\hline
Rapid Rewrite & A write to file $f$ is followed by another write to $f$ within 3 steps that overwrites or reverses it, and the reference does not also rewrite. & Cross-trajectory write sequence and reference alignment. & Edit-stability classification evidence. \\
\hline
\rowcolor{gray!15}\multicolumn{4}{|l|}{\rule[0.0pt]{0pt}{10pt}\textit{Oracle-grounded milestones}}\\
\hline
Milestones & M1 first anchor read, M2 first anchor write, M3 all anchors written, M4 first passing validation, M5 first justified action. & Anchor set plus canonical actions; unreached milestones are right-censored. & Oracle-grounded pace profile. \\
\hline
\end{tabularx}
\end{table*}

\subsection{Divergence Patterns: Cross-Trajectory Diagnosis}
\label{sec:divergence}

Some process patterns are ambiguous without another run for comparison.
\converge handles these cases by asking a reference-scoped question: \emph{how does one agent deviate from a reference process?}
The study design selects the reference trajectory, for example a resolved run for the same task or a contrast that holds either the scaffold or the model fixed.
The compared trajectory is the run being diagnosed.
\converge aligns the two trajectories and classifies the unmatched or reordered regions as named forms of avoidable effort, instability, or drift.

\label{sec:alignment}
\converge begins with a monotonic longest-common-subsequence (LCS) alignment between a reference $R$ and compared trajectory $C$.
Two actions can match only when their action types, targets, and effect labels are compatible.
Compatibility is action-specific.
File actions compare repo-relative paths, while non-file actions use normalized search scope, command intent, or task intent.
Effect labels are matched conservatively: successful workspace effects can match each other, failure-like and non-workspace effects require the same label, and off-anchor actions are treated permissively so uncertain exploration does not inflate divergence.
These rules preserve process differences such as wrong-file work, reversed writes, and failed actions where the reference succeeded, while avoiding over-counting equivalent local choices.

From the matched pairs and unmatched actions, \converge derives coverage, cost, and ordering metrics that summarize omitted reference behavior, added work, and out-of-order matches.
This alignment is deterministic and runs in $O(nm)$.
To check whether classifications depend on the unit-cost LCS choice, we also compare against a weighted Needleman--Wunsch variant.
After alignment, \converge groups unmatched or reordered actions into contiguous \emph{divergence spans}.
Each span is scored by a few observable metrics, such as off-anchor ratio, reversed or rapid rewrites, and dead-end or out-of-order counts.
Each metric has frozen good and warn thresholds.
A value past the warn threshold is high-tier evidence, and a value in the good-to-warn band is medium-tier.

\converge summarizes evidence from divergence spans in three diagnostic layers: file selection, edit stability, and completion behavior.
The file-selection layer asks whether the compared run explored or changed a different part of the repository from the reference.
The edit-stability layer captures unstable rewrites and repeated failed actions, including error-spiral counts from the structural detector catalog.
The completion-behavior layer captures unresolved dead ends and actions that reach the same targets in an inefficient order.
Within each layer, \converge combines pattern counts with evidence tiers and reports a \emph{strong}, \emph{moderate}, or \emph{weak} classification.
These labels describe agreement with the reference in that layer, not absolute run quality.
A strong classification means there is no medium- or high-tier divergence evidence.
A moderate classification means medium-tier evidence, and a weak classification means high-tier evidence or multiple medium-tier patterns.
All classification thresholds are fixed before the reported analyses, and every classification remains relative to the chosen reference.
To assess reference dependence, we repeat the analysis with a per-task most-efficient reference, defined as the resolved candidate reference with the fewest trajectory steps.

\subsection{Task-Relevant Progress Milestones}
\label{sec:milestones}

Anti-patterns and divergence patterns describe what a trajectory contains and how it differs from another run.
Milestones add a temporal view by recording when the agent first makes task-relevant progress.
A milestone is the earliest step at which a specified event occurs with respect to the benchmark's anchor set.
\tool extracts five milestones from each trajectory:
\textbf{M1}~first read of an anchor file,
\textbf{M2}~first write to an anchor file,
\textbf{M3}~all anchor files written,
\textbf{M4}~first passing validation command,
\textbf{M5}~first justified action.
Each milestone yields a (\emph{reached}, \emph{step-to-first-reach}) pair, and trajectories that never reach the event are right-censored and reported separately.
Because milestones require the anchor set, they are oracle-grounded.
They still operate on one trajectory at a time, so the empirical study uses them as a temporal process summary alongside anti-patterns and divergence patterns.

\section{Empirical Setup}
\label{sec:study}
\subsection{Corpora and Collection Protocol}
\label{sec:dataset}

We collect the trajectory corpus from all 500 SWE-Bench Verified~\citep{DBLP:conf/iclr/JimenezYWYPPN24,OpenAI2024SWEBenchVerified} tasks.
Each task is run under five production settings.
We define a setting as a scaffold plus a model:
Claude Code + Opus~4.6, Codex + GPT-5.4, and OpenCode with Opus~4.6, GPT-5.4, and GLM-5.
This yields 2{,}500 trajectories in total.
This design supports two contrast families.
The three OpenCode variants form same-scaffold contrasts where the model changes.
Claude Code vs.\ OpenCode on Opus and Codex vs.\ OpenCode on GPT-5.4 form same-model contrasts where the scaffold changes.
To measure run-to-run variability, we repeat a 50-task stratified random subsample three independent times under each setting, producing 750 additional trajectories.
The sample is grouped by how many of the five settings resolve each task: 20 easy, 20 medium, and 10 hard tasks.
This repeated-run check estimates within-task variance for each process-profile metric and keeps the main results from depending on single-trace attribution.
For cross-benchmark transfer, we also use all 266 SWE-Bench Pro~\citep{DBLP:journals/corr/abs-2509-16941} Python tasks, collecting all 1{,}330 task-setting runs.
We use SWE-Bench Pro to check whether the main associations and scaffold patterns carry to a harder same-family benchmark.
All settings receive the same issue-only prompt and use CLI-default inference parameters.
Each task runs once in an isolated working directory at the task's base commit.
Failed, timed-out, and zero-patch runs are retained rather than filtered.
We use GPT-5.4 as the semantic step labeler over the closed taxonomy in Table~\ref{tab:step-labels}.

\subsection{Baselines and Controls}
\label{sec:baselines}

Because our focus is process diagnosis rather than task solving, we compare against baselines that use the same runs and the same outcome labels.
The \emph{outcome-only} baseline is resolve rate alone.
The \emph{raw-telemetry} baseline uses steps, tokens, durations, failed tool calls, and edit counts.
We also use \emph{same-outcome and same-task controls}, including resolved-vs-resolved pairs and frozen-threshold reuse.
Together, these baselines separate trajectory-specific signal from outcome differences, generic run cost, and task difficulty.

\section{Results}
\label{sec:results}

The empirical study asks what trajectory structure adds beyond success/failure outcomes.
We organize the evaluation from single-run signals to controlled cross-run comparisons.
\textbf{RQ1} asks which anti-patterns act as failure clues, how same-task comparisons localize divergence, and whether the strongest detector signal transfers after thresholds are frozen.
\textbf{RQ2} asks whether deterministic effect labels explain process differences beyond raw telemetry such as steps and failed tool calls.
\textbf{RQ3} asks whether agent settings with similar resolve rates differ in process profiles and task-relevant progress timing.
\textbf{RQ4} asks whether reference-scoped cross-run classifications are stable under reference changes and whether same-model controls reveal scaffold-driven process differences.

\subsection{RQ1: Controlled Failure Evidence}
\label{sec:rq1}

\begin{table}[!t]
\caption{Detector prevalence by outcome and SWE-Bench Verified difficulty. Resolve/Fail report trajectory-level prevalence over all five settings. Easy/Medium/Hard follow the SWE-Bench Verified human time-to-fix buckets ($<$15m, 15--60m, and $>$1h, with Hard merging the 1--4h and $>$4h buckets)~\citep{OpenAI2024SWEBenchVerified}.}
\label{tab:rq1-prevalence}
\centering
\scriptsize
\setlength{\tabcolsep}{2pt}
\begin{tabular*}{\columnwidth}{@{}l@{\extracolsep{\fill}}rrrrr@{}}
\toprule
\textbf{Detector} & \textbf{Resolve} & \textbf{Fail} & \textbf{Easy} & \textbf{Medium} & \textbf{Hard} \\
\midrule
\multicolumn{6}{@{}l}{\emph{Structural (deterministic)}} \\
Search Loop & 41.1 & 56.1 & 34.2 & 49.5 & 67.6 \\
Re-read Churn & 34.0 & 44.7 & 25.5 & 43.0 & 51.6 \\
Tool Oscillation & 5.0 & 8.8 & 3.4 & 6.7 & 13.8 \\
No Formal Tail Validation & 63.2 & 58.2 & 60.1 & 62.9 & 63.1 \\
Unsupported Completion Claim & 7.6 & 9.5 & 8.0 & 8.2 & 8.0 \\
Shell-over-Tool & 15.0 & 14.9 & 12.0 & 16.6 & 18.7 \\
Structured Plan Absence & 9.1 & 13.3 & 7.0 & 11.0 & 19.6 \\
Redundant Search & 9.6 & 11.4 & 7.7 & 11.4 & 12.9 \\
\midrule
\multicolumn{6}{@{}l}{\emph{Semantic (labeler-based)}} \\
Phase Oscillation & 23.4 & 29.6 & 19.7 & 27.0 & 37.8 \\
Semantic Fruitless Exploration & 38.4 & 47.2 & 32.3 & 44.5 & 56.0 \\
\bottomrule
\end{tabular*}
\end{table}

We use detector prevalence as a first pass over the catalog.
Because each task appears under multiple settings, the aggregate comparisons in Table~\ref{tab:rq1-prevalence} are descriptive rather than causal.
This screen identifies search loop as the clearest failure clue because it is common and associated with failed runs.
Re-read churn provides a secondary but less stable clue, while tool oscillation is weaker.
The difficulty columns qualify this interpretation.
Many detectors become more common as tasks get harder, so the resolved--failed gap can reflect residual task difficulty rather than a failure-specific process.
Other rows reinforce the same caution.
Shell-over-tool mainly reflects scaffold convention, and no formal tail validation occurs more often in resolved runs.
The catalog therefore provides deterministic pattern definitions and step-level evidence for inspection, rather than a list of failure predictors.

\begin{figure}[t]
\centering
\includegraphics[width=\columnwidth]{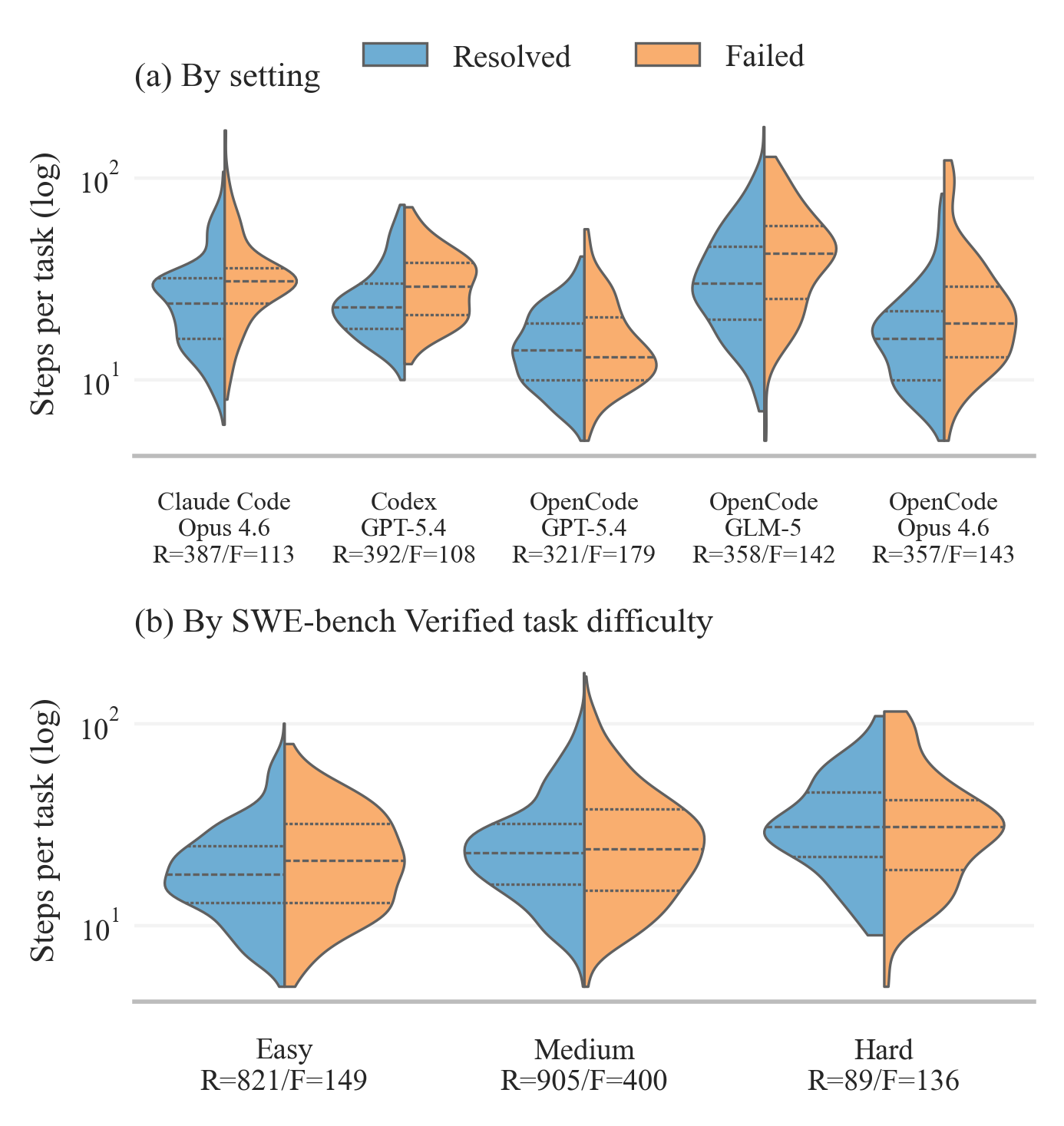}
\caption{Step-count distributions by outcome. Split violins use a log scale, dashed lines mark quartiles, and labels report resolved/failed counts (R/F). Panel (a) groups runs by setting, while panel (b) groups runs by SWE-Bench Verified task difficulty.}
\label{fig:steps-outcome}
\end{figure}

The same limitation appears in a simpler raw-telemetry baseline.
Figure~\ref{fig:steps-outcome} shows that failed runs usually shift toward more steps within each setting, but the resolved and failed distributions still overlap.
Harder SWE-Bench Verified tasks also have longer runs and a higher failure share.
Step count can therefore spot expensive runs at the population level, but it does not explain why a particular run failed.

\begin{table*}[t]
\caption{Same-task divergence by diagnostic layer. Cells report the percentage of comparisons where the failed run departs from the resolved reference. N is the number of task-level comparisons in each row. Function Selection is a finer refinement over comparisons with resolvable function anchors, so its denominator is at most N.}
\label{tab:svf}
\centering
\scriptsize
\setlength{\tabcolsep}{5pt}
\begin{tabular*}{0.95\textwidth}{@{}ll@{\extracolsep{\fill}}rrrrr@{}}
\toprule
\textbf{Contrast} & \textbf{Held fixed} & \textbf{N} & \textbf{File Selection} & \textbf{Edit Stability} & \textbf{\shortstack{Completion Behavior}} & \textbf{Function Selection} \\
\midrule
Claude Code-Opus vs.\ OpenCode-Opus & Model & 52 & 38.5 & 15.4 & 38.5 & 52.4 \\
Codex-GPT vs.\ OpenCode-GPT & Model & 101 & 95.0 & 37.6 & 57.4 & 42.6 \\
OpenCode-Opus vs.\ OpenCode-GPT & Scaffold & 92 & 71.7 & 42.4 & 42.4 & 17.0 \\
OpenCode-Opus vs.\ OpenCode-GLM & Scaffold & 73 & 39.7 & 38.4 & 39.7 & 20.0 \\
OpenCode-GPT vs.\ OpenCode-GLM & Scaffold & 101 & 79.2 & 43.6 & 51.5 & 19.2 \\
\bottomrule
\end{tabular*}
\end{table*}

\noindent\emph{Same-task divergence.}
Detector prevalence and step counts can show that runs differ, but they do not show where the difference occurs.
We therefore align each failed run against a resolved run of the same task.
For each diagnostic layer, we report the percentage of alignments where \converge rates agreement with the reference as moderate or weak.
File selection measures whether the compared run searched, read, or wrote in a different file scope from the reference.
Edit stability measures unstable rewrites and repeated failed actions.
Completion behavior measures unresolved dead ends and inefficient action ordering near the end of the run, while function selection applies the same comparison at the enclosing-function level when anchors are available.
Across the failed/resolved contrasts in Table~\ref{tab:svf}, file selection is usually the largest or tied-largest of the three layers, and completion behavior is also substantial.
The useful signal is not simply that a failed run did more work, but where its work departed from a successful run on the same task.
This points inspection first toward file scope, then toward completion behavior and edit stability.

\begin{figure}[t]
\centering
\includegraphics[width=\columnwidth]{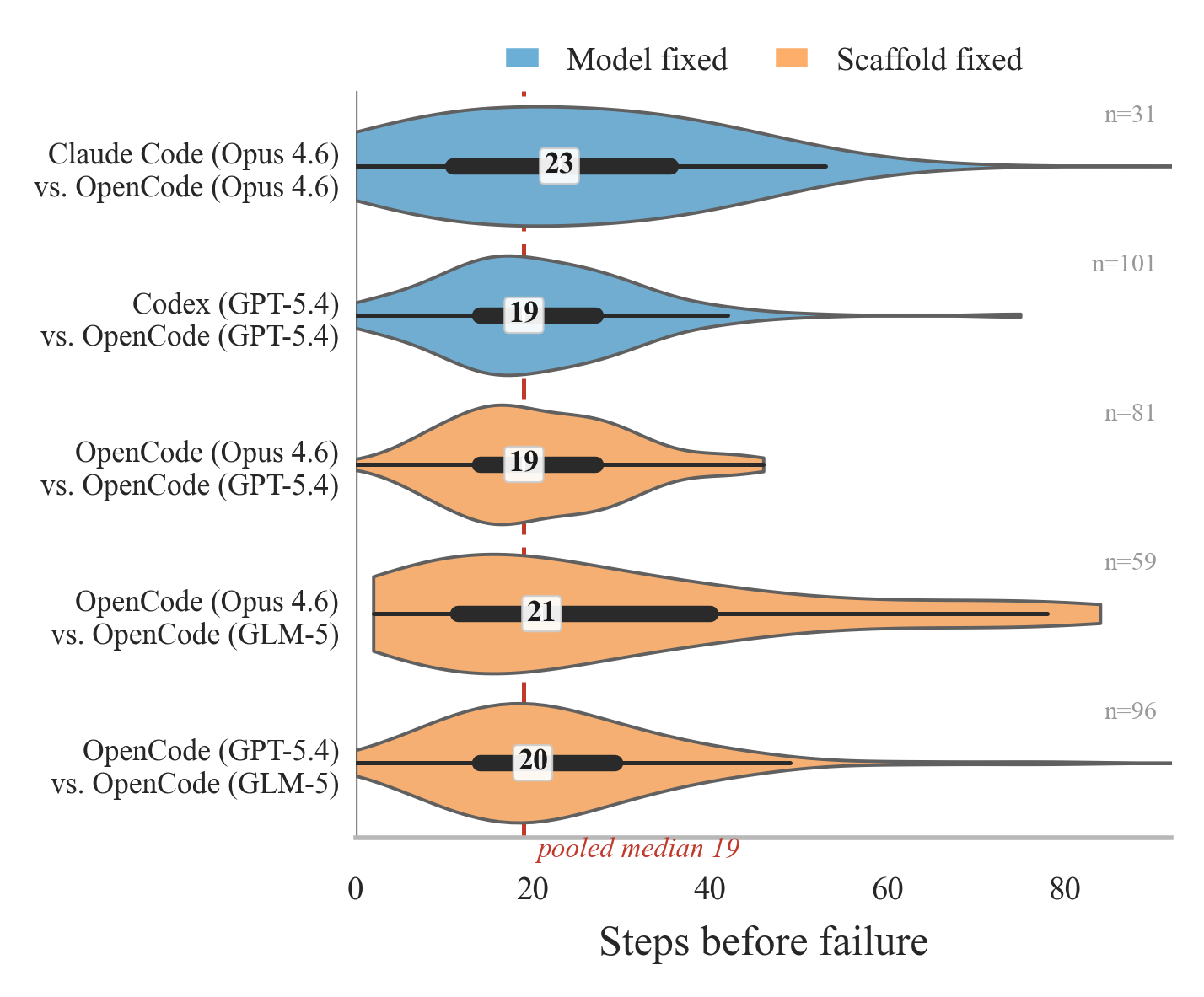}
\caption{Lead time from the earliest same-task function-level divergence to failure. Each violin shows one failed/resolved contrast. Thick black bars mark the interquartile range, labels mark medians, and the red dashed line marks the pooled median.}
\label{fig:leadtime}
\end{figure}

\noindent\emph{Function-level localization.}
The same-task result motivates a finer target view.
File-level divergence is broad because agents can inspect different files before reaching a valid solution.
We therefore resolve each benchmark patch hunk to its enclosing function and ask whether the failed run reached the same code region as a resolved run for the same task.
Figure~\ref{fig:leadtime} shows when the first function-level divergence appears in the failed run.
Across contrasts, the median lead time is around 20 steps before the failed run ends.
This means function-level divergence is available early enough to guide inspection, but it does not prove why the run failed.
The distinction is useful because failures do not all have the same shape.
Some runs work in the wrong file scope, some reach the right file but miss the relevant function, and some reach the right function but still make the wrong edit.
A file-level view collapses these cases, while function-level analysis separates scope recovery from patch reasoning or validation failure.

\begin{table}[t]
\caption{Cross-benchmark transfer check of the frozen structural detectors. The \% columns show how often each detector is observed. Outcome link values below 1 mean the detector is more common in failed runs, while values above 1 mean it is more common in resolved runs.}
\label{tab:pro-transfer}
\centering
\scriptsize
\setlength{\tabcolsep}{2pt}
\begin{tabular*}{\columnwidth}{@{}l@{\extracolsep{\fill}}rrrr@{}}
\toprule
\textbf{Detector} & \textbf{Verified\ \%} & \textbf{\shortstack{Verified\\Outcome Link}} & \textbf{Pro \%} & \textbf{\shortstack{Pro\\Outcome Link}} \\
\midrule
Search Loop & 45.2 & 0.55 & 65.0 & 0.64 \\
Re-read Churn & 37.0 & 0.64 & 30.3 & 0.84 \\
Tool Oscillation & 6.0 & 0.55 & 4.8 & 1.34 \\
No Formal Tail Validation & 61.8 & 1.23 & 46.8 & 0.85 \\
Unsupported Completion Claim & 8.1 & 0.78 & 5.4 & 1.11 \\
Shell-over-Tool & 15.0 & 1.01 & 33.6 & 0.88 \\
Structured Plan Absence & 10.2 & 0.65 & 30.0 & 0.84 \\
Redundant Search & 10.1 & 0.83 & 13.5 & 1.19 \\
\bottomrule
\end{tabular*}
\end{table}

\noindent\emph{Cross-benchmark transfer.}
Finally, we test whether the frozen detector catalog carries beyond SWE-Bench Verified.
We apply it unchanged to all Python tasks in SWE-Bench Pro.
In Table~\ref{tab:pro-transfer}, the prevalence column shows how often a detector is observed.
The outcome link column is directional: values below 1 mean the detector appears more in failed runs, and values above 1 mean it appears more in resolved runs.
Search loop is the most stable transfer signal.
It appears at different rates across the two benchmarks, but its outcome-link value stays below 1 in both, meaning it remains more common in failed runs.
Re-read churn keeps the same direction but moves closer to 1 on Pro, so its association with failure weakens.
Several other detectors move toward 1 or change direction, which means their prevalence is more benchmark-specific.
The useful insight is that detector definitions can transfer, but detector thresholds should be audited on the target benchmark.
Search loop remains a portable warning sign, while the shifted prevalence rates show why practitioners should not reuse thresholds blindly across benchmarks.

\subsection{RQ2: Effect Labels as Diagnostics}

\begin{figure}[t]
\centering
\includegraphics[width=\columnwidth]{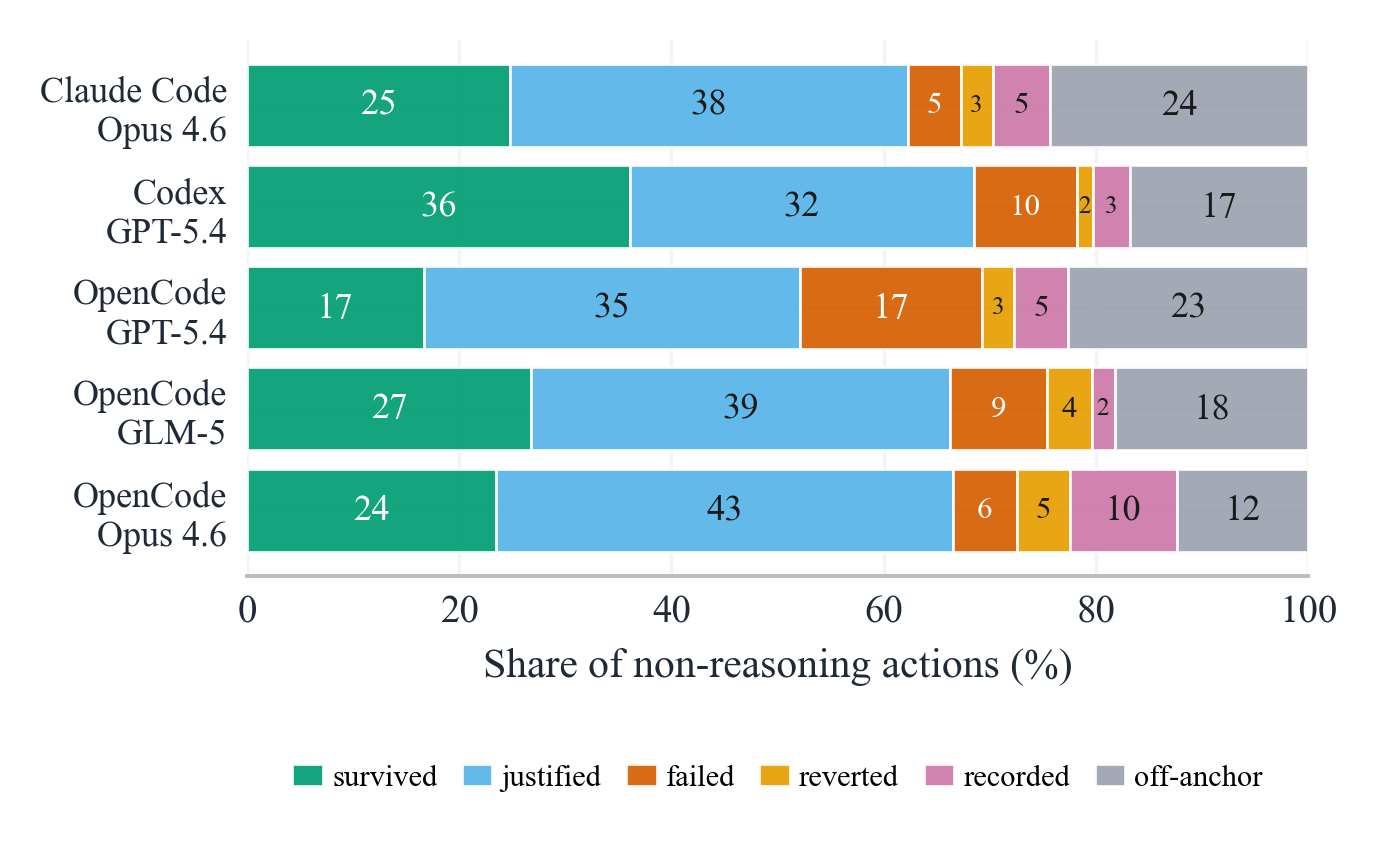}
\caption{Effect-label distribution by agent setting. Each stacked bar shows the share of non-reasoning canonical actions assigned to each effect label. Off-anchor uses the gold-patch anchor set, and the other labels are oracle-free.}
\label{fig:effect-distribution}
\end{figure}

This RQ asks whether deterministic effect labels reveal process differences that raw telemetry cannot show.
Raw counts report how many steps, tool calls, or failures occurred.
Effect labels add a different view: they record whether actions produced surviving changes, failed, were reverted, were justified by task-relevant evidence, or moved outside the anchor set.
We therefore use Figure~\ref{fig:effect-distribution} to compare how agent settings distribute their work across these outcomes.

Effect labels give each setting a compact process profile.
From the figure (per-setting values in Table~\ref{tab:rq3-agents}), Codex has the largest survived-action share, OpenCode/GPT-5.4 has the largest failed-action share, and OpenCode/Opus~4.6 has the lowest off-anchor share.
These differences show that settings can spend their actions in different ways even when raw counts or final outcomes look similar.
The profile yields three practical insights.
The survived-action share shows how much work becomes durable code change, the failed and reverted shares show how much effort is lost, and the off-anchor share shows how much exploration falls outside the benchmark patch scope.
These signals make effect labels useful for comparing settings and selecting runs for inspection.
They should not be read as standalone explanations for why a specific run passed or failed.

\subsection{RQ3: Setting-Level Process Differences}
\label{sec:rq3}

\begin{table}[!t]
\caption{Process profiles and milestone reach by agent setting. Rows from Steps/task through Harmful ratio report per-task medians unless marked as means. Resolve rate and milestone rows report the percentage of tasks reaching each outcome or anchor-grounded event. Off-anchor and milestone rows use the gold-patch anchor set.}
\label{tab:rq3-agents}\label{tab:rq3-milestones}
\centering
\scriptsize
\setlength{\tabcolsep}{3pt}
\renewcommand{\arraystretch}{1.05}
\begin{tabular*}{\columnwidth}{@{}l@{\extracolsep{\fill}}rrrrr@{}}
\toprule
\textbf{Metric} & \multicolumn{1}{c}{\textbf{\shortstack{Claude\\Code}}} & \multicolumn{1}{c}{\textbf{\shortstack{\strut\\Codex}}} & \multicolumn{1}{c}{\textbf{\shortstack{OpenCode\\GLM}}} & \multicolumn{1}{c}{\textbf{\shortstack{OpenCode\\GPT}}} & \multicolumn{1}{c}{\textbf{\shortstack{OpenCode\\Opus}}} \\
\midrule
Resolve rate~\%    & 77.4 & 78.4 & 71.6 & 64.2 & 71.4 \\
Steps/task         & 27 & 24 & 33 & 14 & 17 \\
Tool calls         & 25.0 & 41.0 & 33.0 & 22.0 & 16.5 \\
Tool success~\%    & 96.7 & 91.9 & 100.0 & 100.0 & 100.0 \\
Survived~\%        & 27.6 & 36.2 & 26.1 & 16.7 & 23.1 \\
Failed~\%          & 3.3 & 8.0 & 8.3 & 17.6 & 4.4 \\
Justified~\%       & 36.9 & 32.0 & 37.7 & 34.2 & 41.7 \\
Recorded~\%~(mean) & 5.1 & 4.0 & 2.2 & 4.7 & 9.5 \\
Off-anchor~\%      & 17.6 & 14.3 & 18.1 & 23.0 & 10.2 \\
Harmful ratio      & 0.07 & 0.10 & 0.12 & 0.20 & 0.08 \\
\midrule
\multicolumn{6}{@{}l}{\textit{Oracle-grounded milestone reach (\% of 500 tasks)}}\\
M1 first read      & 99.8 & 98.2 & 99.4 & 99.2 & 97.8 \\
M2 first write     & 95.0 & 95.8 & 93.6 & 91.8 & 94.2 \\
M3 all written     & 85.0 & 86.2 & 83.6 & 82.2 & 84.6 \\
M4 validation      & 2.6 & 2.4 & 4.4 & 1.4 & 2.8 \\
M5 justified       & 99.8 & 99.8 & 99.8 & 100.0 & 100.0 \\
\bottomrule
\end{tabular*}
\end{table}

RQ3 asks whether agent settings with similar resolve rates still behave differently.
Table~\ref{tab:rq3-agents} summarizes each setting with process metrics and oracle-grounded milestone reach.
Most process rows are oracle-free action or effect metrics.
The off-anchor row and milestone rows use the gold-patch anchor set, so they are interpreted only for benchmark tasks with anchors.
Pairwise comparisons are paired by task, and attribution claims are limited to same-scaffold OpenCode or same-model contrasts.

The process rows show why resolve rate alone is incomplete.
Codex has the highest resolve rate and the largest survived-action share, but it also uses many tool calls.
OpenCode/GPT-5.4 has the fewest steps, yet it has the highest failed-action share and harmful ratio, which is the share of a run's non-reasoning actions that failed or were reverted.
OpenCode/Opus~4.6 has the lowest off-anchor share and the highest justified-action share.
These differences show that settings with similar resolve rates can still differ in durable edits, failed work, and task-scope focus.
For practitioners, this profile helps identify what changed after a model or scaffold update, even when the final resolve rate changes little.

\begin{figure}[t]
\centering
\includegraphics[width=\columnwidth]{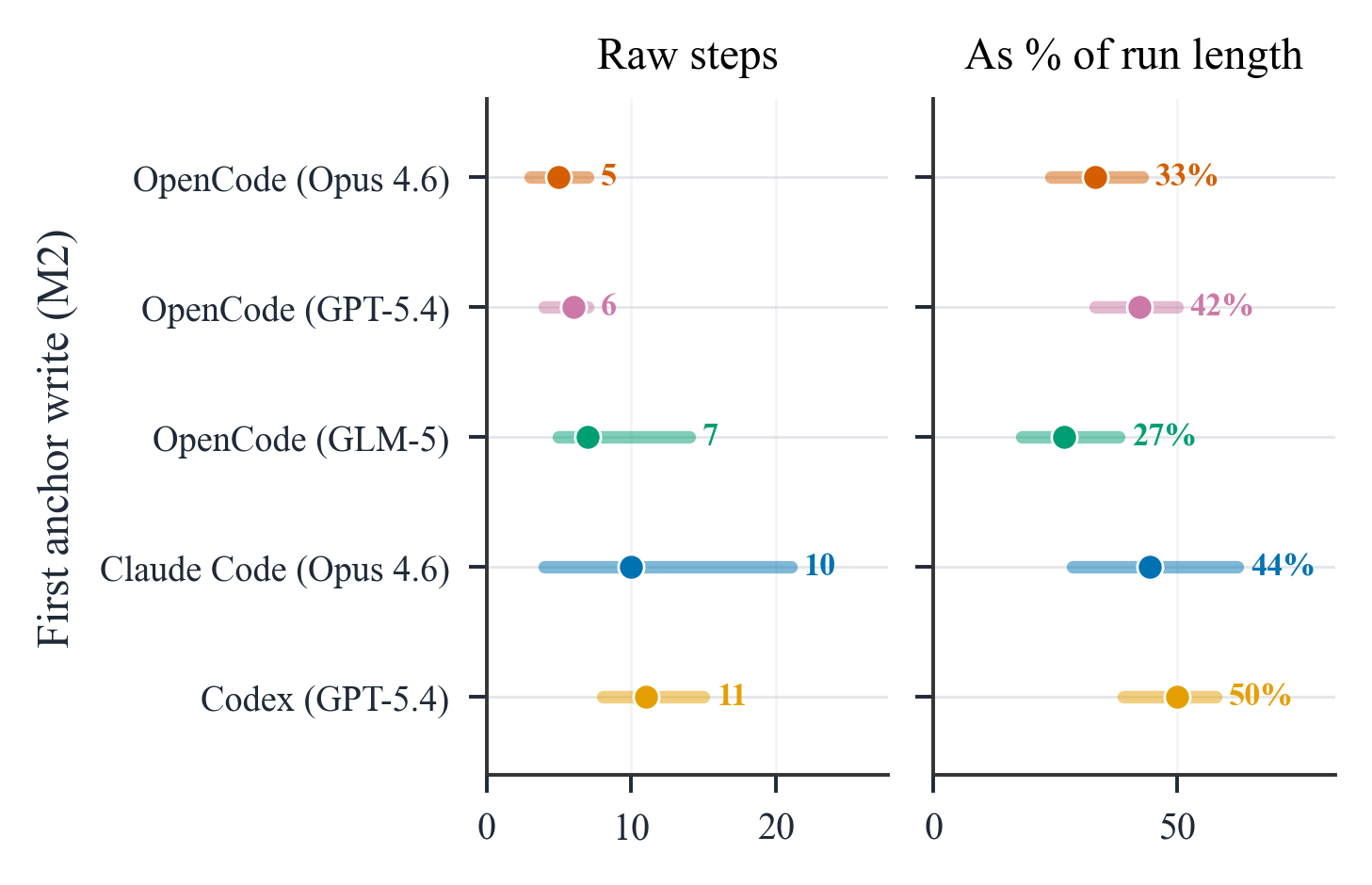}
\Description{A 1x2 row of dot-and-whisker panels of the median step to the first anchor-write milestone (M2) per setting, with IQR whiskers. Columns are raw step count and step normalized as a percentage of run length. The setting ordering changes between the columns: OpenCode/GLM is mid-pack by raw step but earliest as a fraction of its run.}
\caption{First anchor write timing (M2) by setting. Points mark medians and bars mark IQR. The left panel reports raw steps, and the right panel normalizes by run length. Other milestones are omitted because M3 usually coincides with M2 on the single-file tasks that dominate the benchmark, M1 and M5 saturate by step~2, and M4 is reached by under 3\% of runs.}
\label{fig:rq3-milestones}
\end{figure}

Milestones add timing information that the aggregate process rows cannot show (Figure~\ref{fig:rq3-milestones}).
In raw steps, OpenCode/Opus reaches the first anchor write earliest, while Claude Code and Codex reach it later.
After normalizing by run length, OpenCode/GLM moves earlier because it reaches task-relevant files early within longer runs.
This contrast separates quick progress from short runs.
An agent can take more steps overall while still reaching relevant code early in its own trajectory.
Milestones therefore measure progress timing, not patch correctness.

Taken together, the process profiles and milestones give practitioners two complementary checks.
The profile rows show what kind of work a setting performs, while milestones show when the setting reaches task-relevant code.
This makes RQ3 useful for comparing model or scaffold updates even when resolve rate changes little.

\subsection{RQ4: Reference-Scoped Comparisons}
\label{sec:rq4}

\begin{table}[!t]
\caption{Reference-scoped divergence by diagnostic dimension. Cells report the percentage of same-model scaffold comparisons where the compared run has moderate or weak agreement with the fixed reference. Reference change reports how often the agreement category changes after replacing the fixed Codex reference with the per-task most-efficient resolved run. Small reference-change values, especially for file selection, show that these comparisons are largely insensitive to which resolved reference is chosen. The reference-change check covers 435 shared tasks.}
\label{tab:rq4-reference-sensitivity}
\centering
\scriptsize
\setlength{\tabcolsep}{2pt}
\begin{tabular*}{\columnwidth}{@{}l@{\extracolsep{\fill}}rrr@{}}
\toprule
\textbf{\shortstack{Diagnostic\\dimension}} & \textbf{\shortstack{GPT-5.4\\divergence \%}} & \textbf{\shortstack{Reference\\change \%}} & \textbf{\shortstack{Opus 4.6\\divergence \%}} \\
\midrule
File selection & 93.0 & 0.69 & 18.6 \\
Edit stability & 31.4 & 22.1 & 13.2 \\
Completion behavior & 47.0 & 8.7 & 13.6 \\
\bottomrule
\end{tabular*}
\end{table}

RQ4 checks how \converge behaves when the comparison reference changes and when the model is held fixed while the scaffold changes.
The GPT-5.4 contrast compares Codex-GPT with OpenCode-GPT, and the Opus~4.6 contrast compares Claude Code-Opus with OpenCode-Opus.
These same-model contrasts let us ask whether scaffold-driven process differences are visible beyond resolve rate.

The middle column of Table~\ref{tab:rq4-reference-sensitivity} is the reference-change check.
\converge aligns a compared run against a resolved reference run for the same task and assigns strong, moderate, or weak agreement for each diagnostic dimension.
Because a task may have multiple resolved runs, we recompute the comparison with the fixed Codex reference and with the per-task most-efficient resolved reference.
Reference change is the percentage of comparisons where that agreement category changes.
File selection changes in only 0.69\% of comparisons, so file-scope evidence is almost insensitive to reference choice.
Edit stability changes in 22.1\% because recovery and rewrite behavior can vary even among resolved runs.
Completion behavior falls between them at 8.7\%.
Thus, reference-scoped comparison is robust for file selection, while edit stability should be read with more reference sensitivity.

The outer columns show that scaffold effects depend on the model being held fixed.
Under GPT-5.4, the two scaffolds diverge strongly, especially in file selection.
Under Opus~4.6, the same kind of scaffold contrast produces much less divergence in every diagnostic dimension.
This means that a scaffold comparison is not a constant property of the scaffold pair.
Holding the model fixed exposes process differences that endpoint resolve rates can hide.
At the same time, the result is not an absolute quality ranking of the scaffolds; it describes how a compared run departs from a specified reference.

\noindent\emph{Cross-benchmark scaffold control.}
SWE-Bench Pro provides a further same-model scaffold check.
Claude Code and OpenCode share Opus~4.6 and have similar resolve rates on this corpus, yet their detector patterns differ.
Across the 266 shared Pro tasks, structured plan absence is higher for Claude Code (49.6\% vs 0.4\%) and re-read churn is higher for OpenCode (43.6\% vs 16.9\%).
The difference is therefore not simply that one scaffold records more of every detector pattern.
This result reinforces the RQ1 caution that a prevalence shift can be a scaffold signature unless it is checked against outcome, same-task, or same-model evidence.
The Pro scaffold control therefore complements the detector-transfer result.
The same frozen detector signals can reveal both outcome association and process shifts, but the interpretation depends on the contrast being analyzed.

\FloatBarrier

\section{Discussion and Threats to Validity}
\label{sec:discussion}

\noindent\emph{Discussion.}
This study shows why process diagnosis should be reported alongside outcome evaluation.
Two runs can share the same success/failure outcome while following different processes, so a \tool report should be read as inspection evidence rather than automatic remediation advice.
To expose these differences, \insight identifies recurring anti-patterns that help prioritize runs and task families for review.
For cross-run comparison, \converge localizes divergence against a specified reference, while effect labels and milestones describe what work was performed and when task-relevant code was reached.
Taken together, these views provide the strongest evidence through same-task localization at file and function granularity, the stable search-loop clue, and the deterministic representation that makes comparisons repeatable.
At the same time, the controls narrow the claim.
Several aggregate signals weaken under task-fixed analysis, and the transfer check preserves only part of the detector pattern.
For practical use, process profiles should accompany resolve rate.
Practitioners should prefer controlled contrasts over aggregate prevalence shifts, and detector thresholds should be audited on the target benchmark before conclusions are reused.
Under this scope, \tool supports inspection and comparison, but it is not a replacement for resolve-rate evaluation, a per-run causal oracle, or a remediation method.

\noindent\emph{Threats to validity.}
The main calibration threat is benchmark-specific detector tuning.
Several detector thresholds were chosen on SWE-Bench Verified.
The Verified split and raw-threshold sensitivity check assess stability within that corpus after thresholds are fixed, while the SWE-Bench Pro check asks whether the strongest direction carries to a disjoint benchmark.
These checks support stability after thresholds are fixed and bounded transfer, not independent threshold discovery.
However, they do not show that the same thresholds will produce the same prevalence rates or conclusions on every benchmark.
\tool is general because new trajectories can be normalized, labeled, and compared through the same pipeline, but benchmark-specific claims still require benchmark-specific threshold audits.
Beyond calibration, sampling and construct limits also shape the claims.
Each main-corpus task/setting pair is run once, outcome labels inherit SWE-Bench test-suite limits, and \converge agreement assessments measure reference-relative divergence rather than absolute quality or human-validated labels.
Both benchmarks are Python issue to PR repair, so results may not transfer to other languages or task types, and we do not yet evaluate whether the diagnostics change practitioner outcomes.
Execution measurements add another limitation because they can depend on the provider and scaffold.
Duration, token counts, and monetary cost can change with CLI versions, provider-side batching, tokenizer accounting, rate limits, tool permissions, and budget-stop reporting.
For that reason, we interpret them as observed process-cost measurements in this collection, not as intrinsic model speed or efficiency.

\section{Conclusion}
\label{sec:conclusion}
Success/failure outcomes are necessary for evaluating coding agents, but they do not explain the process that produced an accepted or failed run.
This work addresses that gap by treating coding agent trajectories as structured process evidence.
\tool normalizes raw traces into canonical actions and deterministic effect labels, uses \insight to identify recurring single-trajectory anti-patterns, and uses \converge to compare runs against explicit references.
In our SWE-Bench Verified study and SWE-Bench Pro transfer check, \tool identifies population-level failure clues, localizes where failed and resolved runs diverge, and shows that resolved runs can still differ in progress timing, failed work, and observed process cost.
These findings support inspection and comparison, while the controls keep the claims short of explaining exactly why a particular run failed or automatically fixing it.
Future directions include pairing these deterministic diagnostics with human-validated root-cause analysis and intervention studies.
Those studies can test whether changes to prompts, retrieval, memory, tool policy, or validation strategy improve agent behavior while preserving controlled comparisons and keeping deterministic evidence separate from human judgment.

\let\oldthebibliography\thebibliography
\renewcommand{\thebibliography}[1]{%
  \oldthebibliography{#1}%
  \setlength{\itemsep}{-0.5pt}\setlength{\parsep}{0pt}\setlength{\parskip}{0pt}%
  \linespread{0.97}\selectfont}
\bibliographystyle{IEEEtran}
\bibliography{references}

\end{document}